\documentclass[prd,twocolumn,amsmath,amssymb,floatfix,superscriptaddress,nofootinbib]{revtex4}

\usepackage{bm}
\usepackage{amsmath}
\usepackage{epsfig}
\usepackage{color}
\usepackage{natbib}
\usepackage{graphicx} 

\makeatletter
\renewcommand\footnoterule{%
  \vspace{-5pt}
  \kern-3\p@\hrule\@width.4\columnwidth%
  \kern10\p@}
\makeatother

\def\be{\begin{equation}}
\def\ee{\end{equation}}
\def\ba{\begin{eqnarray}}
\def\ea{\end{eqnarray}}

\newcommand{\lsc}{\mathcal{L}}

\newcommand{\lnlmax}{-2\ln\lsc_{\rm ML}}

\newcommand{\kpiv}{k_{\rm pivot}}
\newcommand{\Npiv}{N_{\rm pivot}}
\newcommand{\Npivirh}{N_{\rm pivot}^{(\rm IRH)}}

\newcommand{\Hpiv}{H_{\rm pivot}}
\newcommand{\Trh}{T_{\rm RH}}
\newcommand{\rhorh}{\rho_{\rm RH}}
\newcommand{\thetarh}{\theta_{\rm RH}}
\newcommand{\vend}{V_{\rm end}}
\newcommand{\Mpl}{M_{\rm Pl}}
\newcommand{\thetaemp}{\bm{\theta}_{\rm emp}}
\newcommand{\thetapot}{\bm{\theta}_{V}}

\newcommand{\reffig}[1]{Fig.~\ref{plot:#1}}

\definecolor{darkgreen}{cmyk}{0.85,0.2,1.00,0.2} 
\definecolor{purple}{cmyk}{0.5,1.0,0,0}

\newcommand{\ModeCode}{{\sc ModeCode}}
\newcommand{\MultiNest}{{\sc MultiNest}}
\newcommand{\CAMB}{{\sc CAMB}}
\newcommand{\CosmoMC}{{\sc CosmoMC}}

\begin{document}
\title{Bayesian analysis of inflation: Parameter estimation for single field models}

\author{Michael J. Mortonson}\email{mmortonson@mps.ohio-state.edu}
\affiliation{Center for Cosmology and AstroParticle Physics, The Ohio State University, Columbus, OH 43210, U.S.A.\\ \mbox{}}

\author{Hiranya V. Peiris}\email{h.peiris@ucl.ac.uk}
\affiliation{Institute of Astronomy and Kavli Institute for Cosmology, University of Cambridge, Cambridge CB3 0HA, U.K.}
\affiliation{Department of Physics and Astronomy, University College London, London WC1E 6BT, U.K. \\ \mbox{}}

\author{Richard Easther}\email{richard.easther@yale.edu}
\affiliation{Department of Physics, Yale University, New Haven, CT 06520, U.S.A.}

\date{\today}

\begin{abstract}
\baselineskip 11pt
Future astrophysical datasets promise to strengthen  constraints on 
models of inflation, and extracting these constraints requires methods and 
tools commensurate with the quality of the data.   
In this paper we describe \ModeCode,  a new, publicly
available code that computes the primordial scalar and tensor  power spectra 
for single field inflationary models. 
\ModeCode\ solves the inflationary mode equations numerically, avoiding the 
slow roll approximation. It is interfaced with \CAMB\ and  \CosmoMC\ to 
compute cosmic microwave background angular power spectra and perform 
likelihood analysis and parameter estimation. \ModeCode\ is easily extendable to additional models of inflation, and future updates will include Bayesian model comparison. 
Errors from \ModeCode\ contribute negligibly to the error 
budget for analyses of data from Planck or other next generation experiments. 
We constrain representative single field models ($\phi^n$ with $n=2/3$, 1, 2, 
and 4, natural inflation, and ``hilltop'' inflation) using current data, 
and provide forecasts for Planck.  From current data, we obtain weak but nontrivial limits on 
the post-inflationary physics, which is a significant source of uncertainty 
in the predictions of inflationary models, while we find that Planck will dramatically improve these constraints. In particular, Planck will link the inflationary dynamics with the post-inflationary growth of the horizon, and thus begin  to  probe the ``primordial dark ages'' between TeV and GUT scale energies.  

\end{abstract}

\maketitle

%%%%%%%%%%%%%%%%%%%%%%%%%%%%%%%%%%%%%%%%%
\section{Introduction} \label{sec:intro}

The last two decades have seen a sequence of breakthroughs in the understanding of the physical universe. The detection of cosmic microwave background (CMB) anisotropies by COBE in 1992 \cite{Smoot:1992td}, evidence for dark energy in the distance-luminosity relationship for Type Ia supernovae in 1998 \cite{Perlmutter:1998np,Riess:1998cb}, and the sequence of WMAP data releases \cite{Spergel:2003cb,Spergel:2006hy,Komatsu:2008hk,Komatsu:2010fb}, among others, mark turning points in our ability to constrain --- and falsify --- specific cosmological models.  These advances begin to fulfill the long-standing promise that astrophysical data  can directly probe  the first moments after the Big Bang, while simultaneously constraining models of ultra-high energy physics.  Experiments underway and now being developed guarantee that this progress  will continue well into the future.  In  particular, the Planck satellite \cite{Planck:2006uk} has completed a full survey of the sky, and this data should dramatically improve the constraints  on the free parameters in the so-called concordance cosmology.   

Given the quality of present-day data, the primordial perturbations  are fully described by two numbers (e.g. Ref. \cite{Peiris:2009wp}): the amplitude $A_s$ and tilt $n_s$ of the power spectrum of density (scalar) perturbations.  These quantities form the basis of an {\em inflationary sector\/} in the concordance model, if we postulate that their values can be traced back to a phase of primordial inflation. Conversely,  a complete inflationary model {\em must\/} account for these numbers.   As the data improve, this set of parameters can easily expand to include the properties of any primordial gravitational wave background or departures from Gaussianity. Given the tight agreement between the concordance model and current data, the impact of any new parameters is necessarily subleading. Consequently,  additional parameters needed to describe the primordial perturbations can be regarded as {\em fingerprints\/} of specific inflationary models, in that {\em most\/} of these quantities will be vanishingly small in {\em most\/} inflationary models (see e.g. Ref.~\cite{Baumann:2008aq}). 

Physically, inflation is characterized by a period of accelerated expansion in the early universe, and an inflationary model is defined by the mechanism that drives this accelerated expansion.  Simple models of inflation can usually be described by the kinetic term and potential of a single scalar degree of freedom (the inflaton), along with this field's coupling to gravity. In this paper we focus on models where the field is minimally coupled to gravity and has a canonical kinetic term, but will relax these restrictions in future work.  

The simplest approach to constraining inflation is to specify the primordial perturbations in terms of the empirical parameters
\begin{equation}
\thetaemp = \{A_s,n_s,\alpha_s,...; r,n_t,...; f_{\rm nl},...\},
\label{eq:empparam}
\end{equation}
where $\alpha_s$ is the running of the scalar spectral index, $r$ 
the tensor-to-scalar ratio, $n_t$ the tensor spectral index, and 
$f_{\rm nl}$ parametrizes non-Gaussianity.  Further,  $\thetaemp$ may be extended indefinitely to include higher-order terms in the expansion of the scalar and tensor power spectra and various deviations from Gaussianity. These additional parameters will generally be more 
difficult to measure in any given dataset than the basic empirical 
parameters $A_s$ and $n_s$.  In all cases, inflationary model predictions 
for the values of the empirical parameters are to be compared with the measured values of these parameters. 

A second approach treats the determination of the inflationary mechanism as an inverse problem, and thus reconstructs the inflaton potential from the data \cite{Copeland:1993ie, Copeland:1993jj, Lidsey:1995np, Habib:2004hd, Habib:2005mh}. For instance, {\em slow roll reconstruction\/} \cite{Peiris:2006ug,Peiris:2006sj,Peiris:2008be,Adshead:2008vn} uses a systematic expansion based on the slow roll hierarchy  \cite{Liddle:1994dx,Hoffman:2000ue,Hansen:2001eu,Kinney:2002qn,Easther:2002rw,Liddle:2003py} together with consistency conditions on the duration of inflation, providing a minimally parametric approach to the inverse problem. Variations to this scheme have been widely discussed in the literature  \cite{Leach:2002ar, Leach:2002dw, Leach:2003us, Kinney:2006qm, Lesgourgues:2007gp, Lesgourgues:2007aa, Hamann:2008pb, Kinney:2008wy}.  The inverse problem does not have a unique solution, and typically encodes  basic assumptions regarding the general class of inflationary models under consideration  \cite{Powell:2008bi}. 

Here we adopt a third approach, in which we assume that inflation was driven by a specific model (i.e.\ form of the potential) with one or more free parameters,
\begin{equation}
\thetapot = \{V_1, V_2, ..., V_n; \thetarh\},
\label{eq:vparam}
\end{equation}
where the $V_i$ parametrize the potential, while $\thetarh$
parametrizes the post-inflationary reheating phase, as we 
discuss later.
These parameters replace the empirical inflationary sector, $\thetaemp$. We estimate their values alongside other cosmological parameters, typically using Markov Chain Monte Carlo (MCMC) analysis \cite{Christensen:2000ji,Christensen:2001gj,Knox:2001fz,Kosowsky:2002zt, Lewis:2002ah,Verde:2003ey,Dunkley:2004sv}. Following parameter estimation for several different inflationary models, Bayesian model selection \cite{Jaynes2003, Parkinson:2006ku,Bridges:2006,Liddle:2006,Gordon:2007xm,Feroz:2008} techniques  will allow us to compare the fidelity with which these different models account for observations of the sky.  
 
Working directly with inflationary models, we can compute the power spectrum numerically without recourse to the slow roll approximation. This removes a lingering uncertainty from the analysis of the inflationary parameter space, and allows us to constrain inflationary models with complicated spectra not well described by  $\thetaemp$. While the exact computation of the power spectrum is numerically straightforward and has been used extensively for slow roll-violating potentials previously \cite{Grivell:1996sr, Leach:2000yw, Adams:2001vc, Peiris:2003ff,Martin:2006rs,Covi:2006ci,Ringeval:2007am,Hall:2007qw,Lorenz:2007ze,Bean:2007eh,Lesgourgues:2007aa,Ballesteros:2007,Hamann:2008pb,Mortonson:2009qv}, such precision is only now becoming necessary with the arrival of higher quality data from Planck.

Even when the slow roll approximation is accurate and 
the spectra simple, there are advantages to 
working with $\thetapot$, rather than $\thetaemp$. Empirical quantities such as $A_s$, $n_s$, $r$, etc.\ are computable functions of the free parameters of any specific inflationary model, and thus constraints on $\thetaemp$ can, in principle, be mapped into constraints on $\thetapot$.   However,  we will see that this approach is of limited use in practice, while the opposite approach of inferring constraints on $\thetaemp$ from constraints on $\thetapot$ (given the underlying inflationary prior) is generally reliable.

Finally, $\thetapot$ explicitly accounts for the theoretical uncertainty in inflationary predictions induced by the unknown thermalization history of the post-inflationary universe \cite{Liddle:2003as, Kinney:2005in,Peiris:2006sj,Peiris:2008be,Kuroyanagi:2009br,Martin:2010kz}, in combination with the scale-dependence of the spectral index. This uncertainty is significant even when the running  $\alpha_s \equiv d n_s/d\ln{k}$ is not itself observable.     This apparent paradox arises because when the running is included, $\thetaemp$ requires at least four free parameters to fix the primordial spectra. Conversely,  $\thetapot$ may have only a single free parameter for the inflationary potential and another for the reheating physics, and these numbers can thus be determined with more precision than those in $\thetaemp$ \cite{Adshead:2010mc}. 

This paper has three primary objectives. The first is to introduce \ModeCode, a  plug-in for \CAMB\ and \CosmoMC\ \cite{Lewis:1999bs,Lewis:2002ah}.  \ModeCode\ provides an efficient and robust numerical evaluation of the inflationary perturbation spectrum, and allows the free parameters in the potential to be estimated within an MCMC computation. Secondly, we use this code to generate constraints on representative single field inflationary models using current data, and give forecasts for the constraints that can be expected from Planck.   
Finally, this analysis underlines the importance of assumptions regarding reheating and thermalization in studies of inflation using data from the next generation of astrophysical datasets.  We will see that current data put weak but nontrivial constraints on the reheating history given an explicit inflationary model. Further,  our forecasts suggest that Planck will link 
 the post-inflationary history with the inflationary epoch,  with significant implications for theories of particle physics between Grand Unified Theory (GUT) scales ($\sim 10^{15}$ GeV) and TeV scales.

\ModeCode\ computes both the scalar and tensor perturbation spectra, via the algorithm described in Ref. \cite{Adams:2001vc}.  A number of common inflationary potentials are already included, and new models are straightforward to incorporate  in \ModeCode\ by supplying the potential and its derivatives. 
 Since the CMB angular power spectra and likelihood are already expressed numerically, nothing is lost (other than a relatively small computational overhead) by solving directly for the mode amplitudes, rather than using an analytic approximation. Attention has been given to ensuring that the initial conditions for the background are self-consistent and that the code ``fails gracefully'' for unphysical  parameter combinations, so that such points are excluded from MCMC analyses.  In addition, we accurately compute the endpoint of inflation and the evolution  of the comoving horizon size during inflation, so as to precisely match scales in the inflationary era with scales in the present-day universe.   

In  a follow-up paper, we will  interface \ModeCode\ with \MultiNest\ \cite{Feroz:2008}, allowing us to compute the Bayesian evidence for the models we analyze.   In addition, it is straightforward to extend \ModeCode\ to scenarios with nonminimal kinetic terms \cite{Lorenz:2007ze,Bean:2007eh}, multi-field models, or even non-Gaussianity \cite{Chen:2008wn}.

%%%%%%%%%%%%%%%%%%%%%%%%%%%%%%%%%%%%%%%%%
\section{Methods} \label{sec:methods}

%----------------------------------------
\subsection{Numerical solution} \label{sec:numerical}

In many circumstances, the slow roll approximation provides a sufficiently accurate description of the inflationary power spectra and has the advantage of expressing them as functions of the potential and its derivatives.  However, we want to avoid the slow roll approximation to maintain full generality and accuracy, and instead compute the initial curvature and tensor power spectra numerically given a specific inflaton potential $V(\phi)$  \cite{Grivell:1996sr,Leach:2000yw,Adams:2001vc}. This approach yields exact numerical results for arbitrary inflationary potentials, up to the intrinsic accuracy of first order gravitational perturbation theory.\footnote{Second and higher order mode-mode couplings lead to both non-Gaussianity and loop corrections to the two-point functions. For the models discussed here, these corrections are very small. }

We begin by reviewing the formalism used for the numerical solution of the mode equations.  We describe the scalar perturbations  with the  gauge invariant Mukhanov potential $u$ \cite{mukhanov:1988, sasaki:1986}, which is related to the curvature  perturbation ${\cal R}$: 
\begin{equation}
u = -z {\cal R}\,, 
\end{equation}
where $z\equiv \dot{\phi}/H$, $H$ is the Hubble parameter, and dots denote derivatives with respect to {\em conformal\/} time. 
The Fourier components $u_k$ obey \cite{stewart:1993, mukhanov:1985,mukhanov:1992}
\begin{equation}
\ddot u_k + \left( k^2 - \frac{\ddot z}{z}\right) u_k = 0\,, 
\label{eq:mode}
\end{equation}
where $k$ is the modulus of the wavevector ${\bf k}$.  The power spectrum is 
defined in terms of the two point correlation function
\begin{equation}
\langle {\cal R}_{{\bf k}\vphantom{'}} {\cal R}^*_{{\bf k}'} \rangle = \frac{2\pi^2}{k^3} \Delta_{\cal R}^2(k)\
(2\pi)^3 \delta^{(3)}({\bf k} -{\bf k}'),
\end{equation}
and is related to $u_k$ and $z$ via 
\begin{equation}
\Delta^2_{\cal R}(k) = \frac{k^3}{2\pi^2}\left| \frac{u_k}{z}\right|^2.
\end{equation} 
In terms of the empirical parameters $A_s$, $n_s$, and $\alpha_s$, 
\begin{equation}
\Delta^2_{\cal R}(k) = A_s \left( \frac{k}{\kpiv}\right)^{n_s-1 + \frac{1}{2} \alpha_s \ln(k/\kpiv) + \cdots} \, ,
\end{equation}
where $\kpiv$ denotes the pivot scale at which the power spectrum is normalized.

Equation~(\ref{eq:mode}) depends on the background dynamics through $z$ and its derivatives. Since the logarithm of the scale factor is a natural time coordinate for numerical solutions of the inflationary mode equations, we express the background equations in terms of $\ln a$. Denoting $d/d\ln{a}$ with a prime ($'$) and recalling that $H = d\ln{a}/dt$ by definition, we write the Einstein equation for $H'$ and the Klein-Gordon equation for $\phi$ as follows:
\begin{flalign}
& H' = -\frac{\Mpl^2}{2} (\phi')^2 H \, ,\label{eq:evol1}\\
& \phi'' + \left( \frac{H'}{H} + 3 \right)\phi' + \frac{1}{H^2} \frac{dV}{d\phi} = 0\, ,\label{eq:evol2}
\end{flalign}
where $\Mpl$ is the reduced Planck mass.
Conveniently, our choice of independent variable gives $\phi'=z$, and with the help of these background equations, the mode equation~(\ref{eq:mode}) can be written
\begin{eqnarray}
 u_{k}'' + \left(\frac{H'}{H} + 1 \right)u_{k}' +   \left\{\frac{k^2}{a^2 H^2} \right. - \left[ 2  - 4\,\frac{H'}{H} \frac{\phi''}{\phi'}  \right.  && \nonumber \\ \left. \left. 
  - 2 \left( \frac{H'}{H}\right)^2  - 5 \frac{H'}{H}  - \frac{1}{H^2} \frac{d^2 V}{d\phi^2} \right] \right\}u_k =  0\,, && 
\label{eq:evol3}
\end{eqnarray}
where the term in square brackets is $\ddot z/(za^2H^2)$.

We begin the integration of the background equations when the mode of interest is still deep inside the horizon (i.e. $k\gg 100 aH$). We set the initial field velocity to its slow roll value, solving only the background equations, ensuring that the (small) initial transient in the velocity is damped away.  When the mode is roughly 1/100th of the horizon size we start to evolve the two orthogonal solutions that contribute to $u_k$, and read off the asymptotic value of $|u_k/z|$ when the mode is far outside the horizon and frozen, as explained in Ref. \cite{Adams:2001vc}. 

The usual mode equation for tensor perturbations, 
\begin{equation}
\ddot v_k + \left( k^2 - \frac{\ddot a}{a}\right) v_k = 0\,, 
\label{eq:tensmode}
\end{equation}
becomes 
\begin{equation}
 v_{k}'' + \left(\frac{H'}{H} + 1 \right)v_{k}' + \left[\frac{k^2}{a^2 H^2} - \left( 2  + \frac{H'}{H} \right) \right]v_k = 0 \, 
\label{eq:tensmode2}
\end{equation}
after transforming the independent variable. The primordial tensor power spectrum is  
 \begin{equation}
\Delta^2_t(k) = \frac{4 }{\pi^2} \frac{k^3}{\Mpl^2}\left| \frac{v_k}{a}\right|^2,
\label{eq:tenspower}
\end{equation}
and the asymptotic value of $|v_k/a|$  is again taken from the numerical solutions.

\subsection{Matching and Theoretical Uncertainties }

Ideally, a complete model of the early universe would be predicted by a candidate theory of fundamental  physics.   In that case, we would know the mechanism by which energy is drained from the inflaton to (re)thermalize the universe,  as well as the equation of state and expansion rate of the primordial universe.  Unfortunately, inflationary model building is not mature enough for this to be done on a routine basis and, as a consequence, the unknown expansion history of the post-inflationary universe introduces a theoretical uncertainty into the predictions of inflationary models \cite{Adshead:2010mc}. We must therefore introduce at least one phenomenological parameter $\thetarh$ to account for our ignorance of post-inflationary physics.  For reasons explained below, we work with  $N$, the number of $e$-folds of inflation between moment at which a specified mode leaves the horizon ($k=aH$)  and the end of inflation, defined by  the instant at which $d^2 a(t)/dt^2 =0$. Further, given a specified pivot scale ($\kpiv$) we can then specify a  corresponding $\Npiv$.   For a given inflationary model,  $\Npiv$ is computed from the {\em matching equation}  \cite{Liddle:2003as, Dodelson:2003vq,Alabidi:2005qi}.   
%RJME changes here (including stronger statement about the worth of N)

The matching depends on the growth of the horizon scale following inflation. This is a function of the detailed composition of the primordial universe,
which is not well known, and the matching is thus intrinsically ambiguous.  The two things we know with certainty are that the universe is not thermalized at the end of inflation, and that it must  be thermalized by MeV scales, when primordial nucleosynthesis occurs.\footnote{ Recent evidence points to the existence of a  cosmological neutrino background \cite{Komatsu:2010fb}, which freezes out at temperatures slightly higher than those that apply during nucleosynthesis.} Given that inflation can be a GUT scale phenomenon, the energy ($\sim \rho^{1/4}$) may change by a factor of $10^{18}$ between inflation and nucleosynthesis --- a much larger factor than that between nucleosynthesis and the present day. 

A common assumption is that the universe is effectively matter-dominated ($a(t)\propto t^{2/3}$) as the inflaton oscillates about the minimum of the potential,  thermalizes at some temperature $\Trh$ (with $\rhorh \propto \Trh^4$), and remains radiation-dominated ($a(t)\propto t^{1/2}$) until the matter-radiation  transition at $z\approx 3150$ \cite{Komatsu:2010fb}.  Even for this assumed history, the uncertainty in $\Trh$ allows a wide range in $N$. For example,  a universe which is effectively matter-dominated between energies of $10^{15}$ GeV to $10^{3}$ GeV needs 9 $e$-folds less inflation after a given scale crosses the horizon than one which thermalizes at $10^{15}$ GeV.  In many simple inflationary models, the running $\alpha_s$ is a few times $10^{-4}$, so the resulting uncertainty in the scalar spectral index $n_s$ is $\Delta N \alpha_s \sim 5 \times 10^{-3}$, of the same order as the statistical error expected from Planck \cite{Planck:2006uk,Adshead:2010mc}.  However, far more extreme possibilities for the post-inflationary expansion rate exist, including  kination ($a(t)\propto t^{1/6}$) \cite{Chung:2007vz},  frustrated cosmic string networks ($a(t)\propto t$) \cite{Burgess:2005sb}, or even a short burst of  thermal inflation \cite{Lyth:1995ka}, and taking these scenarios into consideration greatly magnifies the uncertainty in inflationary predictions for a given model.

The truly fundamental variable which specifies the portion of the inflaton potential being traversed as the pivot mode leaves the horizon is simply   $\phi_{\rm pivot}$ --- the value of the inflaton as the $\kpiv$ leaves the horizon.   However, $\phi$ increases in some potentials and decreases in others, while the mapping $\phi \rightarrow \phi+\phi_0$ produces a new potential with identical inflationary  dynamics, so the numerical value of $\phi$ is not informative on its own. Conversely, $N(\phi)$  has the same interpretation in all inflationary models and is a monotonic (and usually simple) function of $\phi$ for a given potential.    For these reasons we take $\Npiv$ rather than $\phi_{\rm pivot}$   as a free parameter. The coupling between $\Npiv$  and the post-inflationary universe simply reflects the physical reality that  the {\em observed\/} inflationary perturbation spectrum is a function of the post-inflationary expansion history, but our choice ensures that the primordial perturbation spectrum is calculated solely in terms of parameters that describe the inflationary epoch which generated it.\footnote{This approach works for scalar perturbations, since modes at astrophysical scales today remain outside the horizon until after nucleosynthesis, at which point the thermal history of the universe is well constrained.  However, primordial gravitational waves seen by direct detection experiments  can enter the horizon during epochs for which the expansion history is not tightly constrained, and for these we need the full transfer function \cite{Boyle:2005se,Easther:2008sx}.}

As we pointed out above, $\Npiv$ depends on the  detailed expansion history of the post-inflationary universe.  However,  there are a huge number of possible combinations of phases in the early universe, and these can have  strongly degenerate predictions for $\Npiv$.\footnote{For example, a long matter-dominated phase leads to the same prediction for $\Npiv$ as a suitable combination of early radiation-domination and a short secondary period of inflation.}  One can imagine  that the unknown expansion history is replicated by an effective barotropic fluid with equation of state $\tilde w$ \cite{Martin:2010kz,Adshead:2010mc}, which is superficially equivalent to regarding $\Npiv$ as a free parameter.  Unfortunately, as explained in Ref. \cite{Adshead:2010mc},  $\tilde w$  is an ambiguous parameter. Given an explicit inflationary potential, we can always determine the moment at which inflation ends, but we cannot compute $\tilde w$ without specifying an energy scale at which the universe has definitely thermalized: the numerical value of $\tilde w$ is a function of this choice.   Admittedly,  $\Npiv$ has an analogous dependence on the choice of $\kpiv$. However,   assuming slow roll, $\Npiv \sim \log{(\kpiv)}/(1-\epsilon)$ so this dependence is usually transparent. Further, for a given combination of datasets, it is possible to determine an optimal choice of $\kpiv$  \cite{Peiris:2006sj},
which is typically close to the geometric mean of the range of scales contributing to the dataset(s). 

We note that there is a correlation between the spectral amplitude at  $\kpiv$ ($A_s$ in the usual $\Lambda$CDM parameter set), and $\Npiv$. As an example, consider $m^2 \phi^2$ inflation: to first order, $\phi_{\rm end}$ and $N(\phi)$ do not depend on $m^2$, while lowering $m^2$ lowers $A_s$ for fixed $\phi$.  Consequently,  $V_{\rm end}$ will also decrease, ensuring that slightly less growth occurs between the end of inflation and some fixed reference point, such as nucleosynthesis or recombination.  Given the precision with which the spectral amplitude is now measured, this effect is small.  Moreover, a similar degeneracy arises with $\tilde w$, since this parameter is also sensitive to changes in the energy density at the end of inflation, if all other parameters are held fixed.

%RJME changes / additions to above paragraphs.

Theoretical considerations put very broad constraints on $\Npiv$. Firstly, in order to ensure that modes do actually reenter the horizon, we need $\tilde w \ge -1/3$, so that $\rho +3p \ge0$ and $\ddot{a} \le 0$.  This is not incompatible with a secondary period of inflation, but does require that the {\em average\/} expansion is not inflationary.   Secondly, for a barotropic fluid, $\rho\ge p$ is required in order to avoid a superluminal sound-speed, so $\tilde w \le 1$.  We are free to estimate inflationary parameters for a narrower range of $\tilde w$ or $\Npiv$, but it is important to recognize that doing so amounts to imposing a theoretical prior on the properties of the post-inflationary universe.

In what follows, we constrain the free parameters of inflationary potentials for two different reheating scenarios,  {\em general reheating\/} (GRH) and {\em instant reheating\/}  (IRH).   In the latter case, we assume that the universe thermalizes instantaneously as inflation ends, and remains thermalized until matter-radiation equality.\footnote{In practice, $\tilde w$ is not exactly 1/3, if the  number of degrees of freedom in the thermal bath is itself a function of temperature.} In contrast, GRH assumes only that  the universe is thermalized by nucleosynthesis scales and that the {\em average\/} expansion is no slower than radiation-dominated, or $\Npiv \le \Npivirh$, the value computed assuming instantaneous thermalization.    This prior is analogous to stipulating that $-1/3 \le \tilde w \le 1/3$, which implicitly rules out a long kination-like phase. Lower values of $\tilde w$ correspond to smaller $\Npiv$, which for the models considered here leads to an increasingly red-tilted spectrum. Current data are thus more sensitive to lower values of $\tilde w$ for the models considered here.

%RJME small changes in this para.

%----------------------------------------
\subsection{MCMC methodology} \label{sec:mcmc}

Our Markov Chain Monte Carlo methodology is based on a modified version of \CosmoMC\ \cite{Lewis:2002ah}, using Metropolis-Hastings sampling for basic parameter estimation, and nested sampling via a \MultiNest\ \cite{Feroz:2008} plug-in for the calculation of Bayesian Evidence (to be presented in a forthcoming publication).   The  free parameters in the potential $V(\phi)$ (plus the one reheating parameter in the GRH case) are varied in the MCMC chains, along with the other cosmological parameters of the concordance model, and any nuisance parameters associated with the datasets. 

For all MCMC analyses of current CMB data, we run $\geq 6$ chains per model/data combination, requiring the Gelman-Rubin \cite{gelman/rubin} criterion on the eigenvalues of the covariance matrix to be $R-1 \lesssim 0.01$ for convergence. For Planck simulation runs we use 4 chains per model which satisfy $R-1\sim 0.1$. While this convergence criterion is not as rigorous as that used in our main analysis, we expect the resulting uncertainty in the estimated variances to be exceeded by the foreground-removal uncertainties \cite{Verde:2005ff} in the large-angle $B$-mode constraint, which are not taken into account in these forecasts.

%----------------------------------------
\subsection{Initial conditions and reheating} \label{sec:code}

We will now describe the implementation of the inflationary initial conditions and the reheating scenarios in \ModeCode. 

{\it Initial conditions:} Inflationary potentials differ in their sensitivity to initial conditions (see Ref. \cite{Bird:2008cp} and references within). Thus, automatically setting self-consistent initial conditions is a nontrivial issue. The initial value of $\phi'$ is set according to the slow roll equations (i.e. the inflaton is assumed to be initially on the slow roll attractor solution). In addition, for a given set of potential parameters, the algorithm must produce a starting field value $\phi_{\rm init}$ which corresponds to a time well before the modes of interest leave the horizon. For the particular models we consider here, the code produces a first guess for $\phi_{\rm init}$ from the field value needed to achieve $N(\phi_{\rm init})=70$ in the slow roll approximation:
\begin{equation}
N(\phi_{\rm init}) = \frac{1}{\Mpl^2} \int_{\phi_{\rm end}}^{\phi_{\rm init}} d\phi \frac{V}{V_{,\phi}}\, ,
\end{equation}
where $\phi_{\rm end}$ is the field value at the end of inflation and $V_{,\phi}\equiv dV/d\phi$. The code then iterates on this initial guess until a self-consistent value of $\phi_{\rm init}$ is found. For some combinations of potential parameters, it is possible that no set of self-consistent initial conditions exists. In such cases we reject the parameter combination in the MCMC analysis by assigning it a very small likelihood.

{\it Reheating:}  \ModeCode\ evolves the inflationary background solution through to the end of inflation, defined by $d^2 a(t)/dt^2 =0$ which corresponds to
\begin{equation}
\epsilon_H \equiv 2 \Mpl^2 \left(\frac{H_{,\phi}}{H}\right)^2 = 1,
\label{eq:epsilon}
\end{equation}
where $\epsilon_H$ is the first Hubble slow roll parameter.\footnote{The code also includes the capability to define the end of inflation as corresponding to a particular $\phi_{\rm end}$, which would be useful for implementing multi-field models. However, with the exception of the test in Sec.~\ref{sec:accuracy}, we do not make use of this feature in the present work.} This calculation yields the number of $e$-folds between the initial conditions and the end of inflation. The matching equation then gives the scale factor at the end of inflation, $a_{\rm end}$. We connect a physical ``pivot'' wavenumber, $k_{\rm pivot}$, to a particular epoch during inflation using
\begin{equation}
k_{\rm pivot} \equiv a_{\rm pivot} H_{\rm pivot} = a_{\rm end} e^{-\Npiv} H_{\rm pivot},
\label{eq:kmatch}
\end{equation}
where $H_{\rm pivot}$ is the Hubble scale corresponding to $k_{\rm pivot}$, which leaves the horizon $\Npiv$ $e$-folds before the end of inflation. In what follows, the pivot scale is set to $k_{\rm pivot}=0.05$ Mpc$^{-1}$.  For a specific inflationary potential  we can compute $\Npivirh$ from the usual matching equation; the post-inflationary expansion is known by hypothesis once we assume instantaneous reheating and
\begin{equation}
\Npivirh = 55.75 - \log{\left[\frac{10^{16} \mbox{GeV}}{V_{\rm pivot}^{1/4}}\right]  } + \log{\left[\frac{V_{\rm pivot}^{1/4}}{V_{\rm end}^{1/4}}\right]}, 
\end{equation}
where the above expression is drawn from Eq.~(20) of Ref. \cite{Adshead:2010mc}, with appropriate substitutions.\footnote{This expression makes no allowance for the changing number of relativistic degrees of freedom as the Universe cools or for dark energy, and it assumes a sharp transition between radiation and matter-dominated expansion. The resulting approximation will not significantly bias our results using present data, but may need to be addressed in the future.}   In the IRH case we compute simply $\Npivirh$ from this expression and the $\Npiv$ is not an independent variable in the chains. 
In the GRH case we  set the prior $20 < \Npiv < \Npivirh$. The lower limit comes from requiring that cosmologically relevant wavenumbers are far outside the horizon when inflation ends. This is tacitly assumed by our computation of the spectrum from the asymptotic mode amplitude in any case, and values of $N$ near this limit are excluded by the data for all the models  we consider here.   The upper limit is enforced by rejecting any step to a model  for which $\Npiv> \Npivirh$. 

%RJME Changes here. Added equation.

% As discussed in Sec.~\ref{sec:numerical}, this implementation of the reheating uncertainty is equivalent to a formulation using an effective barotropic fluid with equation of state $\tilde w$ \cite{Martin:2010kz,Adshead:2010mc}.

% No longer seems to be needed.  

\subsection{Accuracy and timing} \label{sec:accuracy}

To test the accuracy of \ModeCode, we compare its output with the 
analytic solution for primordial perturbations in the ``power law inflation''
model \cite{Lucchin:1984yf}, where the scale factor evolves as $a(t)\propto t^p$ during inflation
and the potential has an exponential form
\begin{equation}
V(\phi) = V_0 \exp\left(\sqrt{\frac{2}{p}}\frac{\phi}{\Mpl}\right).
\label{eq:exponential}
\end{equation}
Power law inflation is one of the very few known models for which the spectrum 
of primordial perturbations can be computed exactly, making it an 
ideal test case for the numerical solution of \ModeCode.

Since the tensor-to-scalar ratio for this model is $r=16/p$, we must 
take a large value of $p$ to avoid violating present upper limits on $r$; 
here we choose $p=60$. For $p>1$, power law inflation does not end via 
slow roll violation [Eq.~(\ref{eq:epsilon})], so we impose an end to the 
inflationary expansion at $\phi_{\rm end}=\Mpl$. 
We additionally assume $\log (V_0/\Mpl^4)=-8.8$ 
and $\Npiv=50$, yielding power spectra that are reasonably 
consistent with observations.

% ****************************************
\begin{figure}[t]
\centerline{\psfig{file=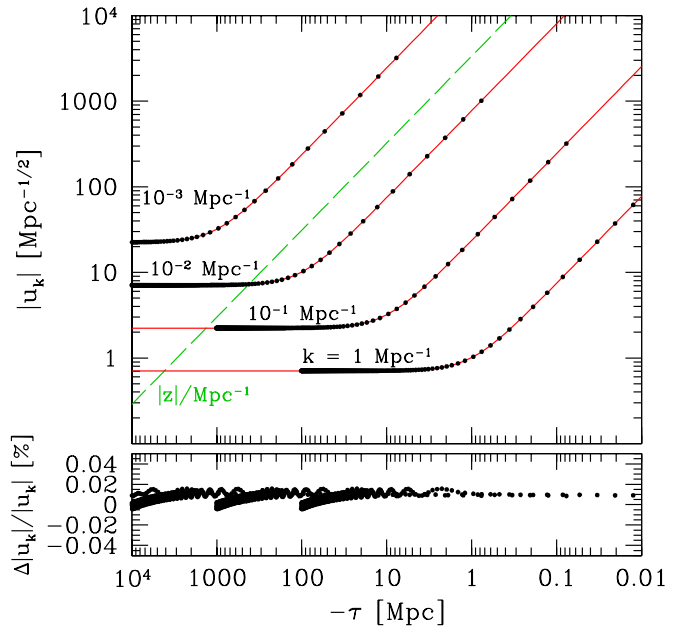, width=3.5in}}
\caption{
Test of the accuracy of \ModeCode\ for power law inflation. Upper panel: 
comparison of the exact analytic solution (solid red curves) and the 
\ModeCode\ solution (black points) for $|u_k|$ as a function of 
conformal time $\tau$ (following Ref.~\cite{Lidsey:1995np}, $\tau$ is 
negative during inflation).
For comparison, the evolution of $|z|$ is plotted as a dashed green line.
Lower panel: percent error in the \ModeCode\ solution for each of the 
points plotted in the upper panel.
}
\label{plot:test}
\end{figure}
% ****************************************

In Fig.~\ref{plot:test}, we compare the \ModeCode\ solution for scalar 
modes with the exact solution from Ref.~\cite{Lidsey:1995np}. 
At all scales, the numerical evolution of scalar perturbations matches 
the exact solution with an accuracy of about $0.01\%$ or better.
This test indicates that \ModeCode\ does not introduce significant 
error in the computation of CMB angular power spectra by \CAMB, which has 
a root mean square accuracy of $\sim 0.3$\% in the configuration to be used for the Planck 
analysis (A. Lewis, private communication).

In the course of previous work using \ModeCode\ 
to compute numerical power spectra for a potential with a step-like feature 
\cite{Mortonson:2009qv}, it was extensively compared with the independent code of 
Ref.~\cite{Adams:2001vc}, yielding agreement to numerical precision. 
It also agrees to similar precision with the mode evolution code used for the same 
potential in Refs.~\cite{Covi:2006ci, Hamann:2007pa}, once the nonstandard choice of 
initial conditions in the latter work is accounted for.\footnote{See Appendix A in 
Ref.~\cite{Mortonson:2009qv} for details.}

To compute the primordial power spectra at arbitrary values of $k$ in \CAMB, 
\ModeCode\ uses cubic spline interpolation on a grid of $k$ values spaced 
evenly in $\ln k$. The extra time required to run \ModeCode\ with \CAMB\ 
depends primarily on the number of $k$ values in this grid. For the default 
setting of 500 grid points over $10^{-5}<k/{\rm Mpc}^{-1}<5$, which 
provides more than sufficient accuracy for smooth primordial power spectra, 
using \ModeCode\ with \CAMB\ typically requires $\lesssim 15\%$ more time 
per evaluation than the default version of \CAMB. The number of grid points 
can be easily adjusted in the code to accurately deal with more complicated 
potentials for which finer sampling in $k$ is required.

%----------------------------------------
\subsection{Data} \label{sec:data}

{\it CMB Data:} We use the $v4$ version of the 7-year WMAP likelihood function with standard options \cite{Larson:2010}, the ACBAR bandpowers from Ref. \cite{ACBAR09} between $550\leq \ell \leq 1950$, and the Pipeline 1 QUaD bandpowers between $569 \leq \ell \leq 2026$ from Ref. \cite{Quad09}. For IRH models, we consider constraints from the 7-year WMAP data (``WMAP7'') only.  In the GRH case, we present both WMAP7 results  and constraints that additionally include data from QUaD and ACBAR (``WMAP7+CMB'').

{\it Planck Simulation:} For selected inflation models in the GRH case, 
we use an unpublished simulation kindly provided by George Efstathiou and Steven Gratton. In this simulation, the model for the ``observed'' power spectra has four components: the primordial CMB power spectra, unresolved point sources, unresolved Sunyaev-Zel'dovich (SZ) clusters, and instrumental noise. The input CMB power spectra are computed from a random realization centered on the ``best fit'' WMAP 5-year cosmological parameters, including $n_s= 0.963$. In addition the simulation includes a tensor component with   $r=0.1$, close to the margin of detectability by Planck \cite{Efstathiou:2009xv,Efstathiou:2009kt}. The likelihood function is described by an exact Wishart distribution, marginalizing over the SZ model and the point source model as nuisance parameters in the MCMC. 
Note that while the fiducial model has $r=0.1$, the particular 
\emph{realization} used in the simulation is consistent with a somewhat 
larger tensor amplitude that is closer to $r=0.14$. The best fit 
values for the models we consider reflect this larger tensor-to-scalar ratio.

\subsection{Models and priors} \label{sec:models}

In order to illustrate our methods, we derive constraints on a variety of models. First we consider a sequence of ``single term'' potentials,
\begin{equation}\label{eq:singleterm}
 V = \lambda  \frac{\phi^{n}}{n} 
\end{equation}
with $n=2/3$, 1, 2, and 4.   The last two cases correspond to the canonical quadratic and quartic chaotic inflation models \cite{Linde:1983gd}, and for  $n=2$ we replace $\lambda$ with $m^2$  in our discussion.  String motivated scenarios \cite{McAllister:2008hb,Flauger:2009ab} can yield potentials with the form  Eq.~(\ref{eq:singleterm}) and non-integer values of $n$ at  large $\phi$. We assume that these potentials are modified for $\phi\le0$ to ensure that $V(\phi) \ge 0$,  if $n$ is not an even integer. 
For convenience, we will work with the simple monomial term; to explicitly constrain  the corresponding stringy scenarios we would instead have to work with the full form of the potential, since the modification near the origin changes $\phi_{\rm end}$ and the matching between $k$ and $\phi$.

For the single term  potentials, the free parameter fixes the height of the potential and the amplitude of the perturbations (i.e.\ $A_s$). However, for these models the other spectral parameters ($n_s$, $r$, etc.) are well approximated by combinations of slow roll parameters, which do not depend directly on $\lambda$.  When the potential has two or more free parameters, the mapping between the explicit form of the potential and the cosmological observables grows more complicated, since these parameters affect both the height and shape of the potential. We give constraints on axion-motivated ``natural inflation'' \cite{Freese:1990rb} with
\begin{equation}
V(\phi) = \Lambda^4 \left[1+\cos\left(\frac{\phi}{f}\right)\right] \, ,
\label{eq:natural}
\end{equation}
and ``hilltop inflation''  \cite{Kinney:1995cc,Kinney:1998md,Easther:2006qu} with
 \begin{equation}
V(\phi) = \Lambda^4 - \frac{\lambda}{4}\phi^4 \ ,
\label{eq:hilltop}
\end{equation}
for which $r$ takes almost arbitrary values while $n_s$ remains close to unity.

% ****************************************
\begin{table}
\caption{Priors on model parameters and maximum likelihood (ML) values for 
WMAP7 GRH constraints. All GRH models include a uniform
prior of $20<\Npiv<\Npivirh$.  Dimensionful quantities are expressed in units where the reduced Planck mass $\Mpl$ is set to unity. Values of $n$ refer to specific cases of  Eq.~(\ref{eq:singleterm}).}
\begin{center}
\begin{tabular*}{\columnwidth}{@{\extracolsep{\fill}}lcllr}
\hline
\hline
Model & Priors & $n_{s,{\rm ML}}$ & $r_{\rm ML}$ & $\lnlmax$ \\
\hline 
$n=2/3$ & $-11<\log\lambda<-7.5$ & 0.965 & 0.07 & 7475.2 \\
$n=1$ & $-11<\log\lambda<-7.5$ & 0.969 & 0.08 & 7475.4 \\
$n=2$ & $-12<\log m^2<-8$ & 0.964 & 0.14 & 7477.3 \\
$n=4$ & $-13.4<\log\lambda<-10.4$ & 0.949 & 0.27 & 7488.7 \\
Natural & $-5<\log\Lambda<0$ & 0.962 & 0.08 & 7475.8 \\
 & $0.5<\log f<2.5$ & & & \\
Hilltop & $-8<\log\Lambda<-2.8$ & 0.944 & $4\times 10^{-5}$ & 7476.2 \\
 & $-13.3<\log\lambda<-12$ & & &  \\
\hline
\hline
\end{tabular*}
\end{center}
\label{tab:priors}
\end{table}
% ****************************************

We show the model priors used in our MCMC analysis in Table~\ref{tab:priors}.  The model parameters correspond to unknown scales in high energy particle physics, so it is natural to sample them logarithmically. For the single parameter models, we will see that the data constrain these scenarios more strongly than the priors on the height of the potential.   On the other hand, the priors chosen for the natural and hilltop inflation models are more restrictive.  Both natural and hilltop inflation have limits in which they are essentially identical to a  $\phi^n$ model, and one of their two free $V(\phi)$ parameters is irrelevant   \cite{Alabidi:2005qi,Adshead:2010mc}.   For natural inflation, this limit is $f \rightarrow \infty$, such that
\begin{equation}\label{eq:natapprox}
V(\phi)\approx \frac{\Lambda^4}{2f^{2}}(\phi-\phi_0)^2
 \end{equation}
 with $\phi_0=\pi f$, which is a quadratic potential after a field redefinition.  In the hilltop case, if $\Lambda$ is very large, the astrophysically relevant portion of the potential is far from the origin. In this limit, 
\begin{equation}
V(\phi)\approx 4\Lambda^4\left(1-\frac{\phi}{\phi_0}\right) \, ,
\end{equation}
where $\phi_0=\sqrt{2}\Lambda/\lambda^{1/4}$ is the field value at which the potential crosses zero. A field redefinition yields a purely linear potential with a single free parameter.    Our priors are chosen to avoid the regions of parameter space where the two parameters of the natural or hilltop inflation model are degenerate.

We stress that the goal of the current paper is to introduce \ModeCode, explore the ability of both current and anticipated datasets to distinguish between different inflationary models, and constrain the number of $e$-folds of inflation required by specific inflationary scenarios.  Broader issues surrounding the choice of priors and model selection will be addressed in a forthcoming paper.

%%%%%%%%%%%%%%%%%%%%%%%%%%%%%%%%%%%%%%%%%
\section{Results}\label{sec:results}

%----------------------------------------
\subsection{Single term potentials} \label{sec:monomial}

% ****************************************
\begin{figure}[t]
\centerline{\psfig{file=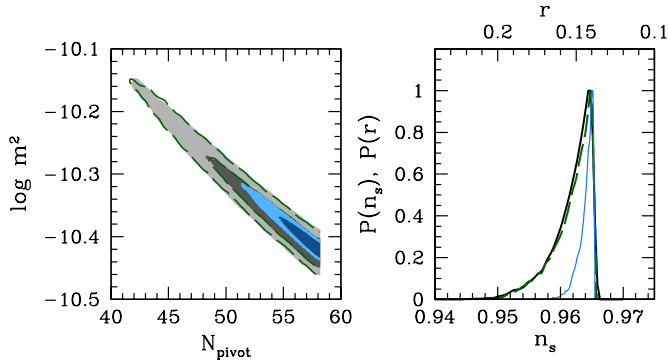, width=3.5in}}
\caption{
Constraints on the quadratic potential ($n=2$). Left: constraints on 
$\log m^2$ and $\Npiv$ for WMAP7 (large gray contours)
showing 68\% CL (light shading) and 95\% CL (dark shading) contours.
Green dashed curves show the contours for WMAP7+CMB data, and the 
small blue contours show simulated Planck constraints.
Right: marginalized 1D $n_s$ and $r$ distributions for 
WMAP7 (thick solid curve, black), WMAP7+CMB (dashed curve, green), and the Planck 
simulation (thin solid curve, blue).
In both panels, the sharp right edges of the distributions correspond to the 
IRH constraints, for which the values of $\Npiv$ (left),
$n_s$, and $r$ (right) are nearly fixed.
Here and in Figs.~\ref{plot:natural} and~\ref{plot:hilltop}, the input parameters for the Planck 
simulation are $n_s=0.963$ and $r=0.1$ (see Sec.~\ref{sec:data}).
}
\label{plot:m2phi2}
\end{figure}
% ****************************************

% ****************************************
\begin{figure}[t]
\centerline{\psfig{file=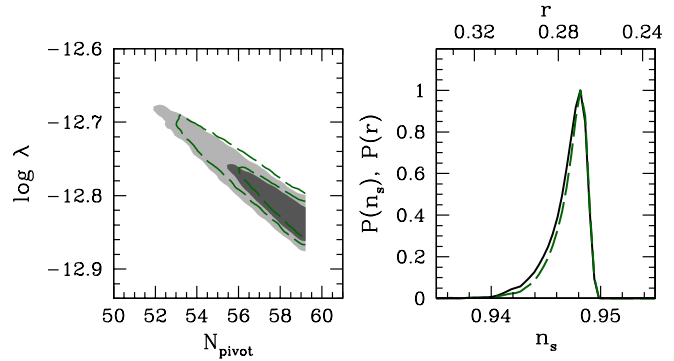, width=3.5in}}
\caption{
Same as \reffig{m2phi2} for the quartic potential ($n=4$).
}
\label{plot:lphi4}
\end{figure}
% ****************************************

% ****************************************
\begin{figure}[t]
\centerline{\psfig{file=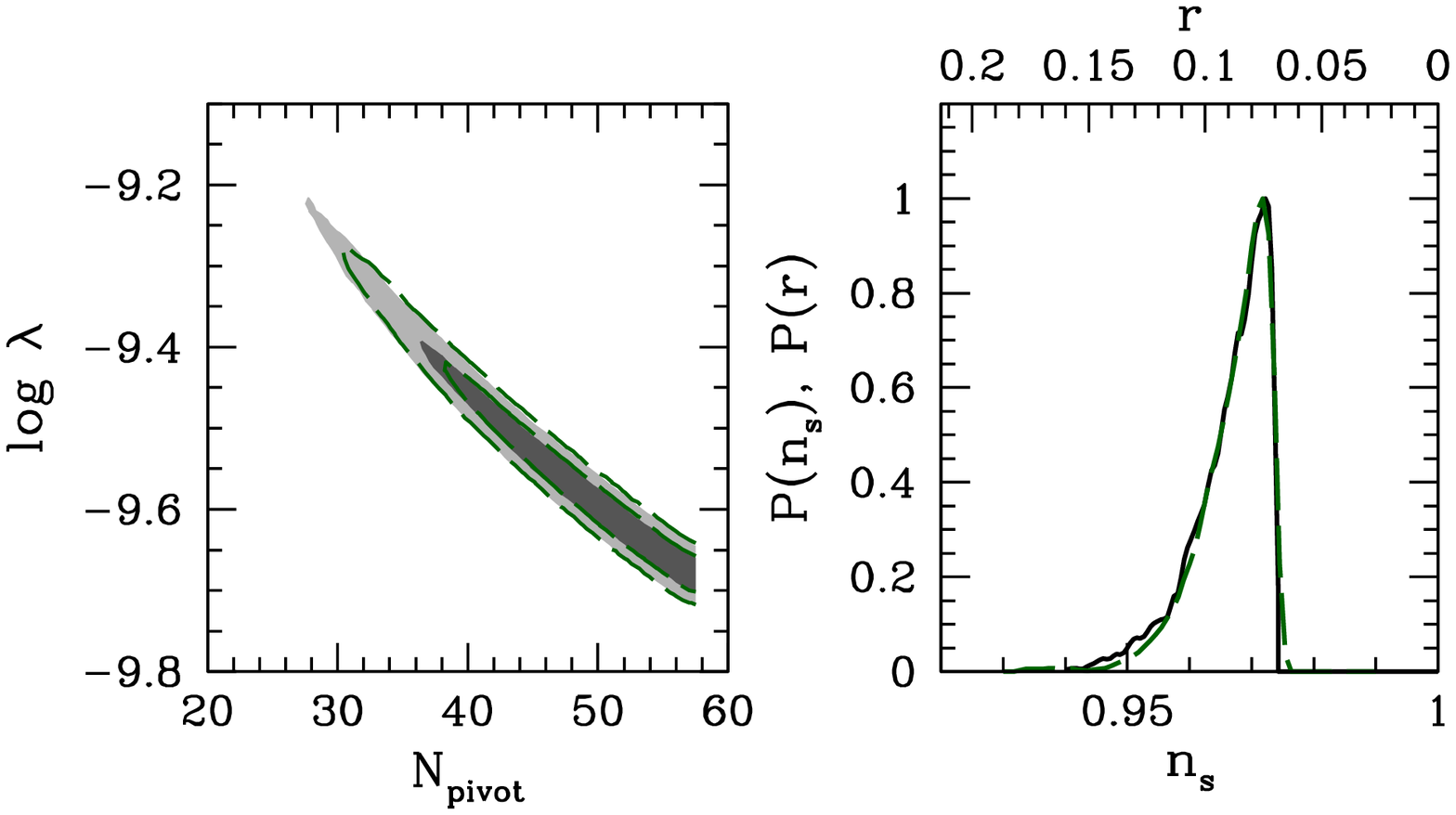, width=3.5in}}
\centerline{\psfig{file=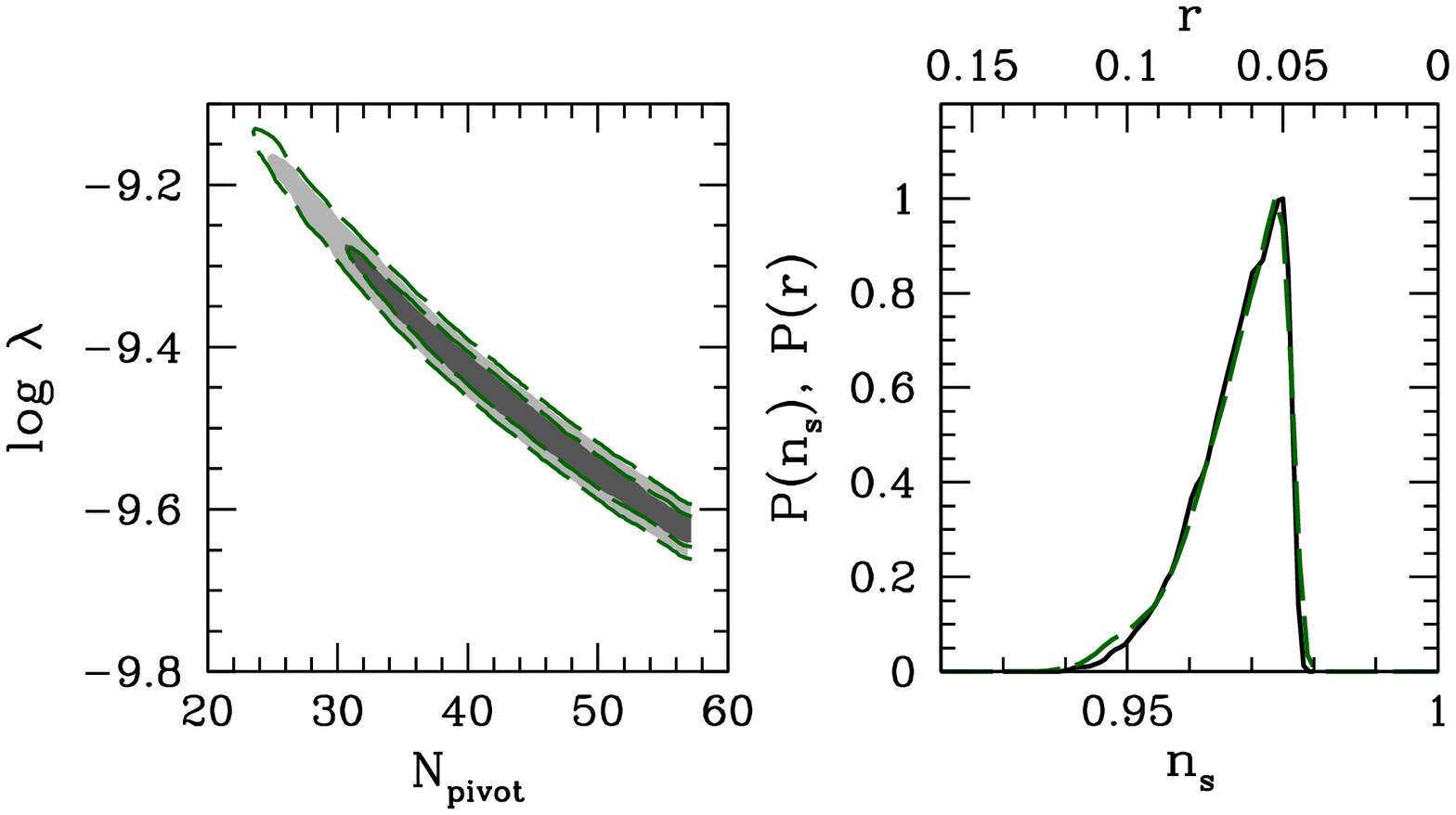, width=3.5in}}
\caption{
Constraints on single term potentials with $n=1$ (top) and $n=2/3$ (bottom).
Conventions match \reffig{m2phi2}.
}
\label{plot:neq2ov3}
\end{figure}
% ****************************************

Figures~\ref{plot:m2phi2}--\ref{plot:neq2ov3} show the 
constraints from WMAP7 and WMAP7+CMB on $\log\lambda$ (or $\log m^2$ for $n=2$) and $\Npiv$ for $V(\phi) =\lambda \phi^n/n$ with $n=2/3$, $1$, $2$ and $4$, along with the inferred constraints on $n_s$ and $r$. 
The limits on $\log \lambda$ from current data are much stronger than the priors listed in Table~\ref{tab:priors}.  Stipulating instant reheating (IRH) ensures that $\Npiv$ is nearly independent of the other model parameters, and only weakly dependent on the value of the exponent $n$: $\Npivirh \sim 57$--$59$. 
In the general reheating (GRH) case, slow roll calculations lead us to expect that $\Npiv$, $n_s$, and $r$ are all effectively functions of a single free parameter, and are thus strongly correlated; these correlations are reflected in the more accurate \ModeCode\ constraints.  The IRH upper limit from the prior on $\Npiv$ creates a sharp cutoff in the distributions for $n_s$ and $r$. 
Given the prior on $\Npiv$, each
of these models has a red tilt ($n_s<1$) and a ``large''  tensor-to-scalar ratio, $r\sim 0.1$.

If $\Npiv$ is lower than its IRH value, or  $\tilde w < 1/3$, $\lambda$ (or $m^2$) is larger than its IRH value: the overall range in this parameter is typically  a factor of $\sim 2$--$3$.  Lower values of $\Npiv$ correspond to smaller $n_s$ and larger $r$, as does increasing the exponent $n$ for fixed $\Npiv$. 
Specifically, the slow roll approximation gives
\begin{equation}
\Npiv+\frac{n}{4} \approx \left(\frac{3}{2}n-\frac{\zeta}{n-1}\right)(1-n_s)^{-1}
\approx \frac{4n}{r}\,,
\end{equation}
where $\zeta=0$ for $n=1$ and $\zeta=1$ otherwise.
Thus upper bounds on $1-n_s$ and $r$ from data can set lower bounds on $\Npiv$.
Except for the $n=4$ case, the predicted value of $r$ is less than the current WMAP upper bound for all values of $\Npiv$ allowed by the prior. For $n=2/3$ and $n=1$ the constraint on $\Npiv$ appears to be driven largely by the correlation between this parameter and $n_s$ ---  in all cases $n_s<0.94$ is strongly excluded. For $n=2$, the larger value of $r$ found with smaller $\Npiv$ provides some additional constraining power. 
Including CMB data on smaller angular scales from QUaD and ACBAR slightly strengthens these limits, relative to the WMAP7 constraints, but does not significantly alter our conclusions.

Of the single term potentials we consider, the quartic $\lambda \phi^4/4$ potential has the largest tensor amplitude and the greatest deviation from scale  invariance.  In agreement with previous analyses of  CMB data (e.g. Ref. \cite{Peiris:2003ff}), this model is excluded by WMAP7 data. Specifically,  $\lnlmax$ exceeds the values found for all of the other models considered here by  $\gtrsim 12$ (see Table~\ref{tab:priors}).  Superficially, this does not appear to be significantly stronger than the result obtained with a single year of WMAP data \cite{Peiris:2003ff}. However, the analysis of   Ref. \cite{Peiris:2003ff}  was carried out at fixed $\Npiv=50$, whereas $\Npiv$ is a free parameter in our chains.  
If we impose the additional prior $\Npiv<50$, the maximum likelihood of the 
quadratic potential worsens to $\lnlmax=7499.4$. This is $\sim 24$ larger than
the overall best fit and excludes such models with much greater confidence than the first year of WMAP data alone. 

In recent work, Martin and Ringeval \cite{Martin:2010kz} quote constraints on the reheating temperature following inflation driven by a $V(\phi) \sim\phi^n$ potential.  The tightest constraints they present imply that the reheat temperature is above the TeV scale. However, this specific constraint is obtained for a prior that renders the post-inflationary expansion rate a function of $n$.   Since scenarios for which $V(\phi) \sim\phi^n$ at large field values can have very different shapes near the origin, the prior could only be realized by a carefully tuned potential, as $V(\phi)$ would need to be well approximated by $\phi^n$ at energies far below the inflationary scale. Moreover, this form of $V(\phi)$ must be modified near the origin if $V(\phi) \ge 0$ for all $\phi$, and the potential does not have a discontinuous first derivative at $\phi=0$ for $n\ne2$ or 4.

%RJME Change here. 
 
Constraints using the simulated Planck likelihood show that the uncertainties in parameters of these single term potentials will be greatly reduced by the next generation of cosmological datasets.  For the quadratic potential, we see from Fig.~\ref{plot:m2phi2} that the simulated Planck constraints strongly disfavor (at $>95\%$ CL) models with $n_s\lesssim 0.96$, corresponding to  $\Npiv\lesssim 50$ and $\log m^2\gtrsim -10.3$.   Recall that for quadratic inflation followed by a matter-dominated phase and then thermal inflation, $\Npiv$ is at least 10 less than the instant reheating value $\Npivirh$ \cite{Lyth:1995ka}. Consequently, we predict that Planck can differentiate between these two post-inflationary scenarios for quadratic inflation.

% ****************************************
\begin{figure}[tbp]
\centerline{\psfig{file=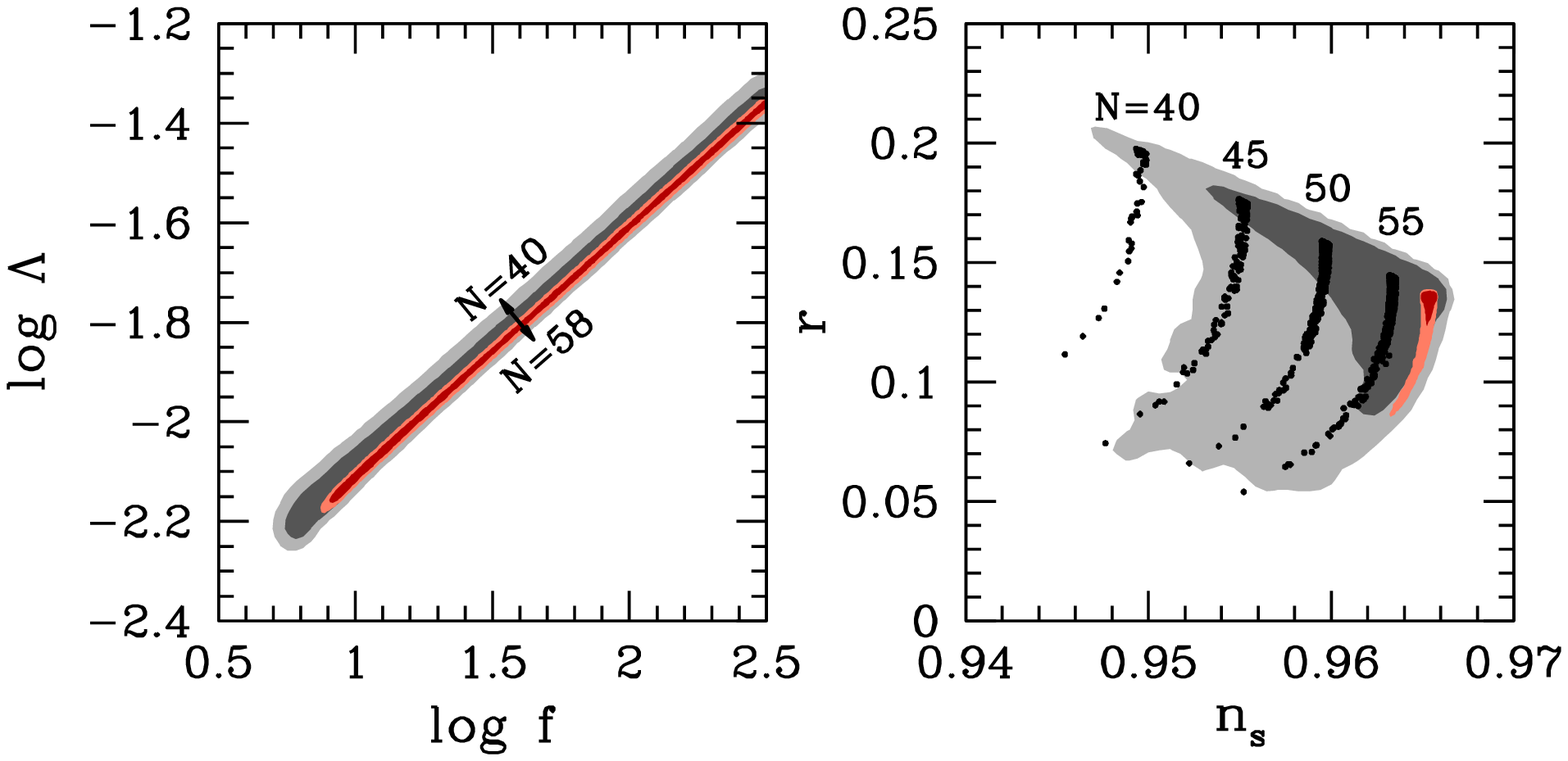, width=3.5in}}
\vspace{0.5cm}
\centerline{\psfig{file=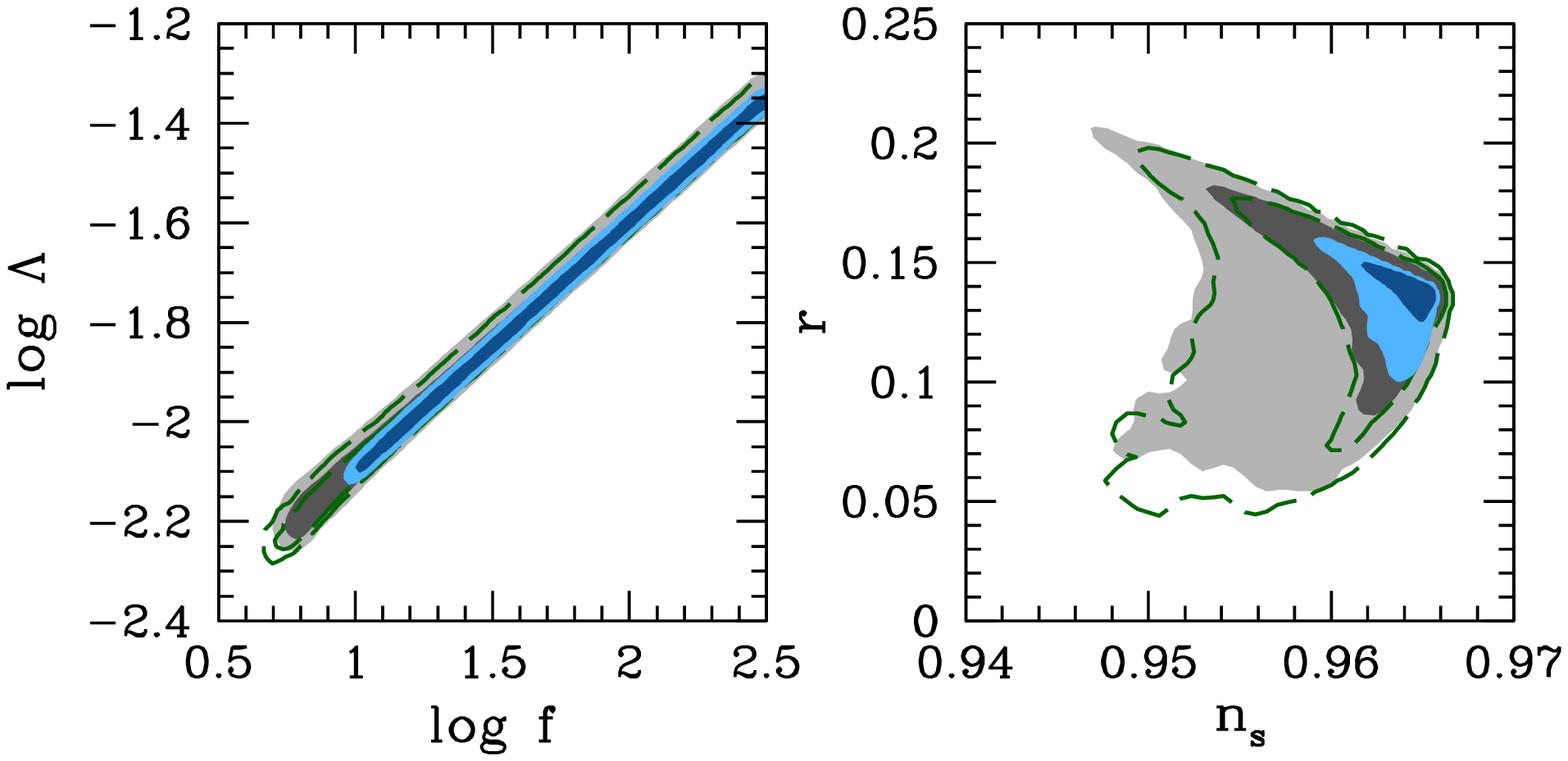, width=3.5in}}
\caption{
Top left: WMAP7 68\% and 95\% CL constraints on natural inflation parameters
$\log \Lambda$ and $\log f$ for the GRH (large contours, gray shading) and 
IRH (small contours, red shading) scenarios.
Top right: constraints on $n_s$ and $r$; points show random samples of 
models from the MCMC analysis for which $\Npiv$ is within 0.25 of the values 
indicated in the plot.
Bottom: Predicted natural inflation GRH constraints from Planck (small contours, 
blue shading) compared with current constraints from WMAP7 
(gray shading) and WMAP7+CMB (dark green, dashed contours).
}
\label{plot:natural}
\end{figure}
% ****************************************

%----------------------------------------
\subsection{Natural inflation} \label{sec:natural}

Figure~\ref{plot:natural} shows our constraints on the natural inflation [Eq.~(\ref{eq:natural})] parameter space and the derived empirical parameters $n_s$ and $r$ from \ModeCode.
The relationship between the empirical parameters and the potential parameters for natural inflation is discussed in detail in Ref. \cite{Savage:2006tr}, along with parameter constraints derived from the 3-year WMAP dataset.

Unlike the single term potentials, current data permit a wide range of natural inflation parameters and, as noted in Sec.~\ref{sec:models}, there is a degeneracy between $f$ and $\Lambda$ in the limit where these parameters are large.   In this region of parameter space, natural inflation overlaps with the quadratic model. Our priors are chosen to exclude most of this region; given the parametrization of the potential, if we allowed arbitrarily large values of $f$ and $\Lambda$ (and given that the quadratic potential is not currently excluded by data) almost all points drawn by the chains would be in this degenerate region.    
Our adopted priors on $\log f$ and $\log \Lambda$ still allow a region of 
nearly-degenerate models that contribute to the ``ridge'' seen in the right panels of Fig.~\ref{plot:natural}; these models closely match the values of $n_s$ and $r$ seen in the quadratic potential constraints.  The marginalized constraints on $n_s$ and $r$ depend strongly on the prior on $\log f$ due to the projection of a large number of degenerate models into this ridge. Thus, the apparent preference for this region of parameter space over models with lower values of $r$ is largely due to this effect, and is not driven by the data.

Instant reheating requires $\Npivirh\sim 58$, similar to the 
constraint for the single term potentials, but the additional 
inflationary degree of freedom in the potential permits a larger 
range of $n_s$ and $r$. More generally, for fixed $\Npiv$, 
decreasing $\Lambda$ and $f$ reduces both $n_s$ and $r$. This is
shown by the MCMC samples with fixed $\Npiv$ plotted in the upper right panel
of Fig.~\ref{plot:natural}. Thus natural inflation models can have lower values of $r$ than 
the quadratic potential in \reffig{m2phi2},  without increasing $n_s$ and $\Npiv$.

The lower panels of Fig.~\ref{plot:natural} show how the natural inflation 
constraints improve with additional CMB data. As for the single term 
potentials, the difference between WMAP7 and WMAP7+CMB constraints is small, but Planck is expected to yield a dramatic improvement.  
In particular, the uncertainty in $\Npiv$ --- which is visible in the width of the $\log \Lambda-\log f$ contours --- is substantially reduced in the Planck forecast.
Since the Planck simulation we use is not inconsistent with $m^2 \phi^2$ inflation, 
the limit in which natural inflation becomes indistinguishable from quadratic inflation would not be excluded in this particular forecast.  
Conversely, if   quadratic inflation is disfavored by future data, we will be able to put data-driven upper bounds on $f$ and $\Lambda$ in addition to tightening the existing lower bound.

%----------------------------------------
\subsection{Hilltop inflation} \label{sec:hilltop}

\begin{figure}[tbp]
\centerline{\psfig{file=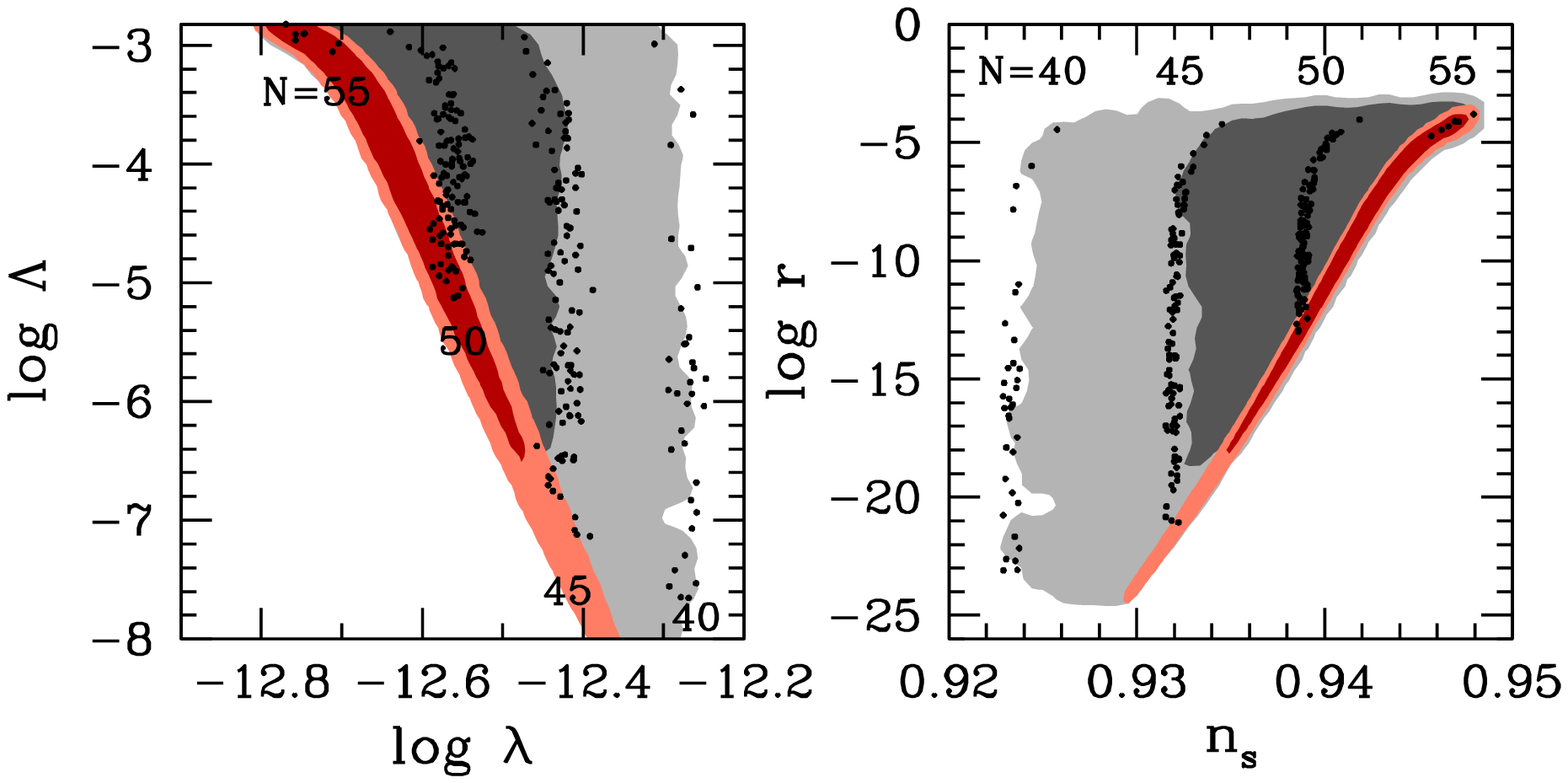, width=3.5in}}
\vspace{0.5cm}
\centerline{\psfig{file=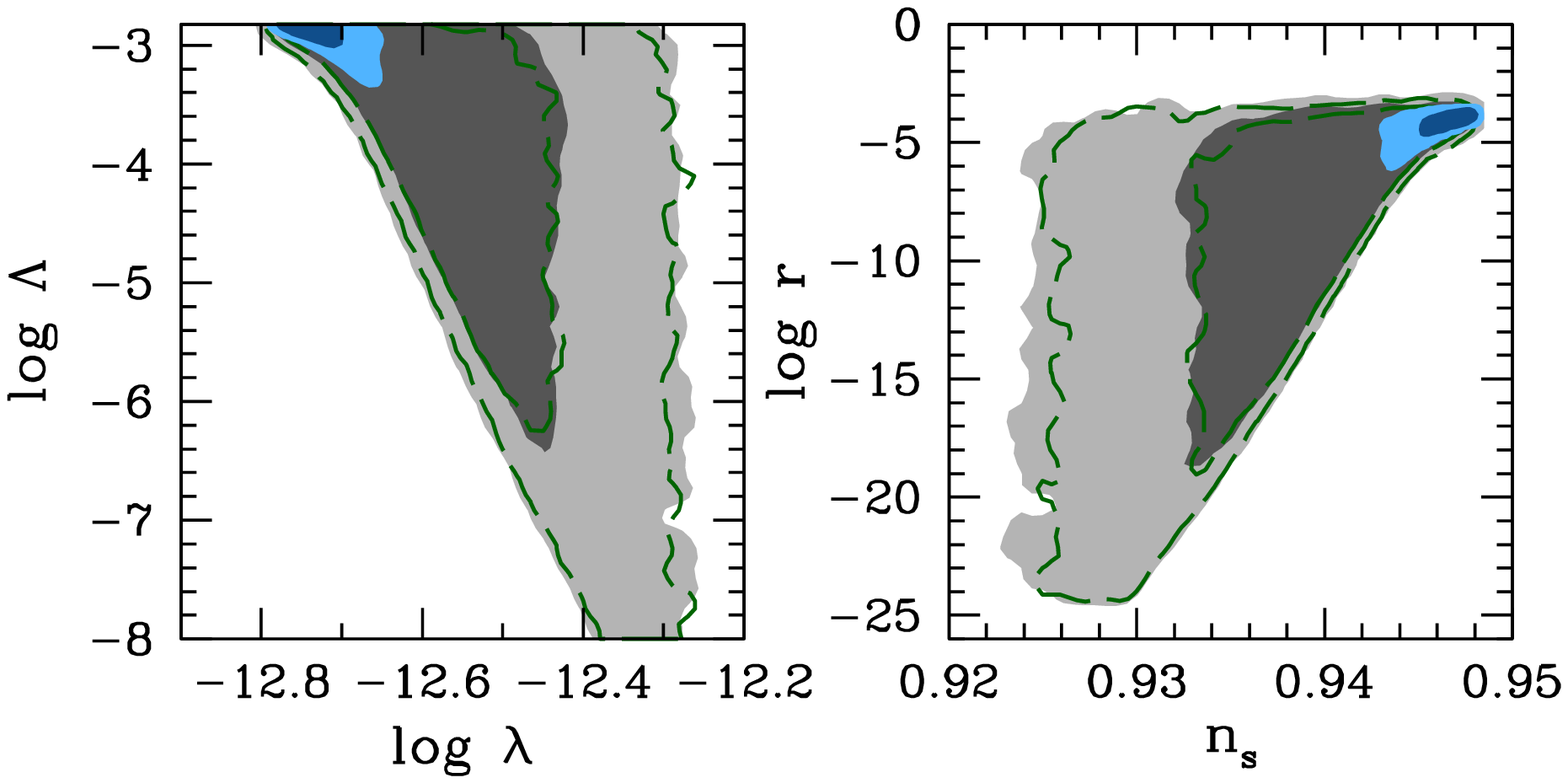, width=3.5in}}
\caption{
Same as \reffig{natural} for the hilltop inflation model with 
parameters $\log \Lambda$ and $\log \lambda$.
Note the logarithmic scale for $r$ in the right panels; the prior here
permits very small values of $r$.
}
\label{plot:hilltop}
\end{figure}

Figure~\ref{plot:hilltop} shows the constraints on $\log \Lambda$, $\log \lambda$, 
$n_s$, and $\log r$ for hilltop inflation. We impose an upper limit of $\Lambda<0.0015$ 
to remove models where the field starts far from $\phi=0$ and near the $V=0$ crossing point,
as described in Sec.~\ref{sec:models}.  The remaining models have a small tensor 
amplitude and relatively large deviations from scale invariance.
The MCMC samples plotted at fixed $\Npiv$ in Fig.~\ref{plot:hilltop}
show that the number of $e$-folds is correlated with $n_s$, but not $r$, over most of 
the allowed region of parameter space.

% ****************************************
% ****************************************

Constraints on the hilltop inflation model from WMAP7+CMB data and 
from the Planck simulation are compared with the (GRH) WMAP7 constraints 
in the lower panels of Fig.~\ref{plot:hilltop}. The values of $n_s$ and $r$ allowed 
by the hilltop model within our chosen priors are smaller than 
those assumed in the Planck simulation, so the contours for Planck 
are concentrated at the largest allowed values of both parameters. 
In fact, for this particular forecast the entire region of 
the hilltop inflation parameter space within our priors would 
be strongly excluded by Planck, with $\lnlmax$ $\sim 75$ larger than its value
for the quadratic and natural inflation models.

% ****************************************
\begin{figure*}[t]
\centerline{\psfig{file=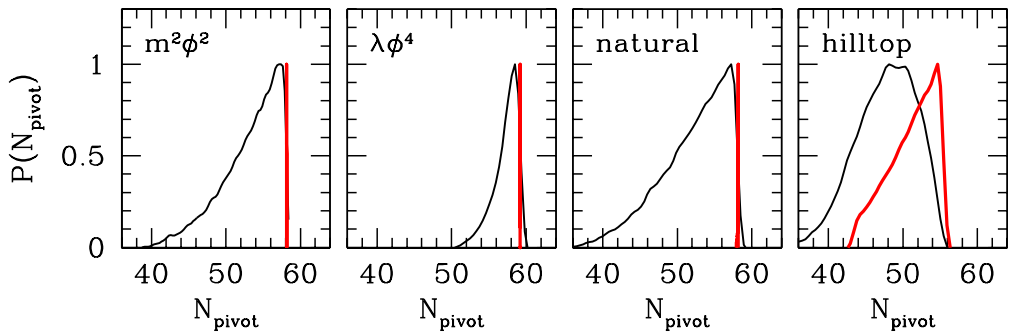, width=5.5in}}
\caption{
Marginalized 1D distributions for $\Npiv$ from WMAP7 in the GRH case 
(thin black curves) and IRH case (thick red curves). From left to right, 
the models are the quadratic potential, the quartic potential, 
natural inflation, and hilltop inflation.
}
\label{plot:npivot}
\end{figure*}
% ****************************************

Unlike the other models considered here, current data allow $\Npivirh$ to cover a substantial range (roughly 10 $e$-folds), as illustrated in  Fig.~\ref{plot:npivot}.  The weak  constraint on $\Npivirh$ is due to a special cancellation in the slow roll expression for the scalar spectral amplitude, which leaves $A_s$ independent of the overall height of the potential.\footnote{In the more general class of hilltop potentials $V(\phi) = \Lambda^4 -\lambda \phi^n /n$,  this situation only occurs for $n=4$  \cite{Kinney:1995cc,Kinney:1998md,Easther:2006qu}, so this property is not generic.}
Meanwhile, the departure from scale invariance is $1-n_s\approx 3/\Npivirh$ 
in the limit of small $\Lambda$.   
Other two-parameter models can formally support inflation at low energy scales, but the spectra of such models are typically far from scale-invariant and thus disfavored by the data.  
On the other hand, hilltop inflation can  have $\Lambda \ll 10^{16}$ GeV, which ensures that the tensor amplitude is very low, without driving the spectral index to an observationally excluded value. Consequently,
both $r$ and $\Npivirh$ can vary greatly, as we see in Fig.~\ref{plot:hilltop}.  Models with very low values of $\Lambda$ do have a lower likelihood, as the stronger breaking of scale invariance  in these models is at odds with the spectral tilt allowed by the data.  For IRH models  there is effectively only one free parameter in the potential after fixing $A_s$, leading to a strong correlation between $n_s$ and $r$ which is absent in the GRH case.

% ****************************************
\begin{figure}[t]
\centerline{\psfig{file=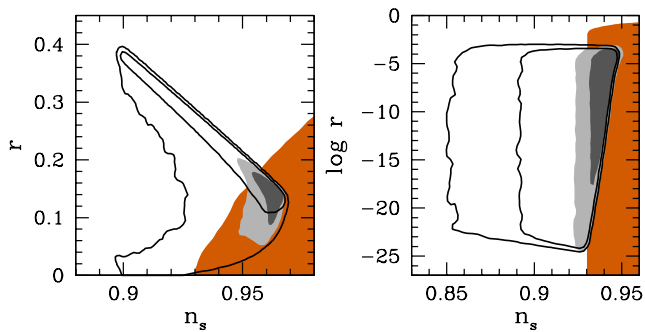, width=3.5in}}
\caption{
The parameter space in $n_s$ and $r$ corresponding to uniform sampling 
within the priors on potential parameters listed in 
Table~\ref{tab:priors} (unshaded curves, 68\% and 95\% CL regions) 
for natural inflation (left) and hilltop inflation (right).
The WMAP7 GRH constraints from Figs.~\ref{plot:natural} and~\ref{plot:hilltop}
are shown again for comparison as shaded gray 68\% and 95\% CL regions.
Orange shading shows the 95\% CL region for empirical WMAP7 constraints 
with flat priors on $n_s$ and $r$.
}
\label{plot:priors}
\end{figure}
% ****************************************

%----------------------------------------------------
\subsection{Impact of model priors} \label{sec:priors}

For the single term potentials, the priors on the amplitude of the 
potential $\lambda$ (or $m^2$) listed 
in Table~\ref{tab:priors} are sufficiently weak compared to the 
constraints from the data that they have no significant impact on 
the estimated parameter values and confidence regions. 
The upper limit on $\Npiv$ corresponding to instant reheating, however, 
does significantly reduce the allowed region in parameter space
for each of these models, thus restricting the possible values of 
$n_s$ and $r$.

The resulting upper limit on $n_s$ and lower limit on $r$
can lead to tension with observations. For example, 
the WMAP7 constraints are $n_s=0.982^{+0.020}_{-0.019}$ (68\% CL) and 
$r<0.36$ (95\% CL) \cite{Larson:2010}, treating these as empirical parameters without
specifying a particular potential. For the quartic $\lambda\phi^4/4$ potential, 
the instant reheating limit on $\Npiv$ and the nearly perfect correlation 
between $\Npiv$, $n_s$, and $r$ result in the limits $n_s\lesssim 0.95$
and $r\gtrsim 0.27$; since both of these are in tension with the 
measured values, the quartic potential provides a poor fit to the 
data relative to the other single term potentials which can 
achieve larger $n_s$ and smaller $r$.

On the other hand, as noted earlier the natural inflation and hilltop 
inflation priors have been chosen specifically to limit the 
extent of parameter degeneracies that would otherwise be allowed by 
current data. In the limiting regions that are truncated by the priors, 
these two models are degenerate with specific monomial potentials,
as discussed previously.  Furthermore, the mapping between $\thetapot$ and $\thetaemp$
for these models is significantly more complicated than for the 
single term potentials, so the uniform top-hat priors on $\thetapot$
can correspond to highly non-uniform prior distributions for $\thetaemp$. This mapping of the priors is illustrated in 
Fig.~\ref{plot:priors}, which shows the regions
of the $(n_s,r)$ plane obtained by uniform sampling within the 
priors on $\thetapot$ specified in Table~\ref{tab:priors} for the 
natural inflation and hilltop inflation models.

Due to the natural inflation degeneracy between $\log \Lambda$ and $\log f$, even in 
the absence of any data the priors 
clearly favor models along the line $r\approx 4(1-n_s)$ corresponding to 
the approximately quadratic regime near the minimum of the natural 
inflation potential. The WMAP7 GRH constraints in 
Fig.~\ref{plot:natural} are qualitatively 
described by the intersection of the priors 
with the empirical constraints on $n_s$ and $r$ from WMAP7.
Note that the upper limit on $n_s$ is typically set by the $\Npiv\leq\Npivirh$
prior (the lower right edge of the 95\% CL region in the left panel 
of Fig.~\ref{plot:priors}), while the lower limit on $n_s$ --- and thus $\Npiv$ --- follows from the data.
The 95\% lower limit on $r$ does not quite reach down to the limit allowed by the prior because 
of marginalization over the strong projection effect described in Sec.~\ref{sec:natural}, which favors
models along the upper diagonal ridge of the prior.
With present data, the region in the $(n_s,r)$ plane allowed by the priors is 
fully consistent with the measured values of these parameters (and 
therefore the best fit $n_s$ and $r$ values in Table~\ref{tab:priors} 
are consistent with the WMAP7 empirical constraints), but upcoming 
measurements may yet rule out the natural inflation model.

For the hilltop inflation model, the constraints on $n_s$ and $r$  
are partially influenced by the priors on $\log \Lambda$ and $\log \lambda$ 
from Table~\ref{tab:priors}; in particular, the range of $r$ allowed is 
limited by the prior on $\log \Lambda$ (see Fig.~\ref{plot:priors}). 
However, the distribution of models allowed by the priors in the $(n_s,\log r)$
plane is much more uniform for hilltop inflation than it is for 
natural inflation. As for natural inflation, the upper limit on $n_s$ 
for hilltop inflation models is set by the $\Npiv\leq\Npivirh$ prior, 
and the lower limit on $n_s$ is determined by the data.
Like the quartic potential, the upper limit on $n_s$ corresponding to 
instant reheating is low compared with the preferred value from WMAP7;
however, in the case of hilltop inflation, 
this is coupled with a small value of $r$, which enables the hilltop 
inflation model to fit the WMAP7 data reasonably well since the 
empirical constraints on $n_s$ and $r$ are correlated.

\subsection{Slow roll mapping}

Many previous constraints on inflationary models have been obtained by taking the empirical parameters $\thetaemp$ (e.g.\ $A_s$, $n_s$, and $r$) predicted by a given model and  comparing these with constraints on empirical parameters (see e.g. \cite{Martin:2006rs,Komatsu:2010fb}).  Given current data, limits on $A_s$, $n_s$ and $r$ typically dominate the constraints on simple inflationary models; starting from constraints on these empirical parameters allows constraints on a number of inflationary models to be inferred from a single set of empirical Markov chains.

% ****************************************
\begin{figure}[t]
\centerline{\psfig{file=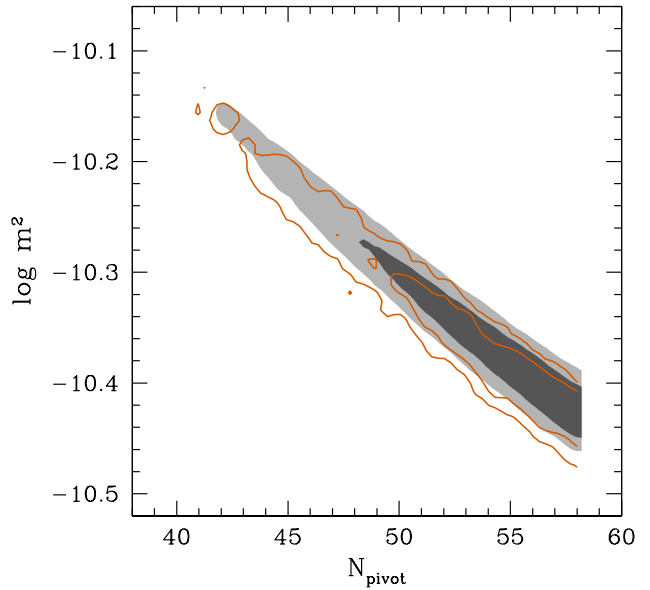, width=3.5in}}
\caption{
Approximate constraints on the quadratic potential
obtained by selecting MCMC samples from standard WMAP7 $\Lambda$CDM+tensor 
chains and mapping $A_s$, $n_s$, and $r$ to $\log m^2$ and $\Npiv$ using 
the slow roll approximation. The shaded gray contours are 
the WMAP7 GRH constraints from \reffig{m2phi2}, obtained by a 
direct MCMC estimate of $m^2$ and $\Npiv$ for quadratic inflation. 
The unshaded orange contours are derived by sampling chains for $A_s$, $n_s$, 
and $r$, from which we selected only those points with  $|1-n_s-r/4|<0.005$ 
to approximately match the slow roll relation between $n_s$ and $r$ 
for the quadratic potential.
}
\label{plot:m2phi2map}
\end{figure}
% ****************************************

For the simplest inflationary models, this approach allows one to quickly constrain many potentials at once. In general, however, the mapping $\thetaemp\to\thetapot$  is not one-to-one; thus this method requires sampling within the space of potential parameters and is no more efficient than directly constraining these parameters from the data using \ModeCode. For any potential parametrized by $\thetapot$, one can compute the primordial scalar and tensor power spectra (as described in Sec.~\ref{sec:numerical}, or otherwise) and find the values of 
$\thetaemp$ by  fitting to these power spectra. However,  for any given inflationary potential many values of  $\thetaemp$ cannot  be obtained for any  combination of $\thetapot$. Likewise, as we saw for the natural and hilltop scenarios, some models have degenerate combinations of $\thetapot$ which correspond to the same primordial power spectra, so the mapping $\thetapot\to\thetaemp$ is not  always invertible.  Further, models with sharp features or other complexities do not produce a power spectrum that is easily  described by the usual empirical parametrization.

Despite these problems, it is instructive to attempt to derive constraints on $\thetapot$ from  constraints on $\thetaemp$ for the purposes of comparison with our main results in the previous sections.  
Here we perform these tests using the slow roll approximation, although one could implement the mapping using more precise methods. In Appendix~\ref{sec:appendixA} we describe the slow roll mapping for the quadratic potential. 
Figure~\ref{plot:m2phi2map} shows that the constraints on $\log m^2$ and $\Npiv$ obtained by mapping from empirical parameters agree reasonably well with the direct MCMC constraints from \ModeCode. However, they are visibly noisy due to the smaller number of MCMC samples, and are systematically shifted toward lower values of $\log m^2$, due to the mapping being done at lowest  order in slow roll. 

While this approach works well for the quadratic potential (and can be expected to produce similar results for the $n=1$ and $n=2/3$ cases)  it is far less efficient for  the quartic ($n=4$) potential. Recall that this scenario is disfavored by the data. Consequently,  using the Metropolis-Hastings algorithm  to draw samples from the $(n_s,r)$ parameter space results in very few (if any) accepted points in the relevant region for the quartic potential on the $(n_s,r)$ plane, and thus the contours of $\lambda$ and $\Npiv$ are extremely noisy. Constraining the quartic potential parameters using this method requires MCMC sampling that is specifically designed to acquire samples in the $(n_s,r)$ region spanned by this model.

Models with multiple parameters like natural inflation and hilltop inflation face an additional problem in that the mapping from 
$\thetapot$ to $\thetaemp$ is not invertible in regions where there is a parameter degeneracy. For example, at fixed values of $\Lambda^4 f^{-2}$, all natural inflation models with large $f$ have nearly identical power spectra (see Sec.~\ref{sec:models}). Therefore, a single point in $\thetaemp$ space cannot be 
simply mapped to the corresponding values of $\thetapot$ in this 
degenerate region. Such a mapping would require an additional MCMC run or 
some similar method of sampling in the $\thetapot$ parameter space.

%%%%%%%%%%%%%%%%%%%%%%%%%%%%%%%%%%%%%%%%%
\section{Discussion} \label{sec:discussion}

In this paper we introduce \ModeCode, a new, publicly available numerical solver for the inflationary perturbation equations. \ModeCode\ is extendable, numerically efficient, and integrated with \CAMB\ and \CosmoMC.   We demonstrate the use of \ModeCode\ by  constraining several single-field inflationary models using current CMB data, confirming that present-day data put useful constraints on the parameters of a number of interesting and well-motivated inflationary models. 

Using a simulated likelihood, we present forecasts for the quality of the constraints that can be expected from Planck, showing that these will greatly strengthen limits on the inflationary parameter space and possibly exclude some simple inflationary models. 
In the Planck forecast analysis, our aim was not to provide ``Fisher''-style forecasts for each model in turn, but to take the realization provided by the simulation and analyze it with our pipeline as we would the real sky. Given the level of tensors present in this simulation, which is at the margin of detectability with Planck, we find that small-field models would be excluded with high significance as expected. Conversely, in a scenario where Planck polarization measurements did not find evidence of tensor fluctuations, we predict that many large-field models 
would be either excluded or limited to a small region of parameter space. 

In our analysis we pay close attention to the interplay between the post-inflationary expansion history and the inflationary observables.  These are connected by the matching criterion, which determines  the moment during the inflationary epoch at which a given comoving scale leaves the horizon. Inflation is often assumed to be a GUT scale phenomenon, but the expansion rate and thermal state of the post-inflationary universe is not {\em directly\/} constrained until MeV scales, at which point the success of Big Bang nucleosynthesis and evidence for a cosmological neutrino background strongly suggest that the universe was thermalized.   Consequently,  there is a huge range of energies over which the composition and expansion rate of the universe are effectively undetermined.    
 
For most  inflationary models, the  constraints on the total number of $e$-folds since the pivot scale left the horizon, $\Npiv$, are noticeably tighter than the hard lower bound ($\Npiv\ge20$) assumed in our analysis.  Thus we can be confident that our constraints on $\Npiv$ are driven by the data, although these currently eliminate only   relatively extreme post-inflationary scenarios.    By contrast, the Planck forecast suggests that the next generation of CMB data will put much tighter constraints on the reheating history, given a specific inflationary model. For instance, for $m^2 \phi^2$ inflation, Planck should discriminate between a long matter-dominated phase that extends to the TeV-scale and instant reheating at perhaps the $2~\sigma$ level.   

This correlation between the post-inflationary dynamics and the inflationary epoch has significant consequences for  particle physics. 
For instance, many supersymmetric scenarios predict that the primordial universe undergoes a period of matter domination driven by heavy moduli (e.g.\ \cite{Easther:2008sx,Acharya:2010af}), for which $\Npiv$ differs substantially from the instant reheating value, and post-Planck cosmology will thus be increasingly concerned with the full  evolutionary history of the universe. Moreover, inflationary model builders will be able to definitively test scenarios which include predictions for the post-inflationary expansion rate.

\section*{Acknowledgments}
We are extremely grateful to George Efstathiou and Steven Gratton for the permission to use their unpublished Planck simulation, which was used to generate the ``Planck forecast" contours in Figures~\ref{plot:m2phi2},~\ref{plot:natural}, and~\ref{plot:hilltop}. MJM is supported by CCAPP at Ohio State. HVP is supported in part by Marie Curie grant MIRG-CT-2007-203314 from the European Commission, and by STFC and the Leverhulme Trust. RE is partially supported by the United States Department of Energy (DE-FG02-92ER-40704) and the National Science Foundation (CAREER-PHY-0747868).  HVP and RE thank the Aspen Center for Physics for hospitality during the completion of part of this work. RE thanks IoA Cambridge and DAMTP CTC for additional hospitality. Numerical computations were performed using the Darwin Supercomputer of the University of Cambridge High Performance Computing Service (http://www.hpc.cam.ac.uk/), provided by Dell Inc.~using Strategic Research Infrastructure Funding from the Higher Education Funding Council for England. We acknowledge the use of the Legacy Archive for Microwave Background Data (LAMBDA). Support for LAMBDA is provided by the NASA Office of Space Science.

\appendix

%----------------------------------------
\section{Slow roll mapping---quadratic potential} \label{sec:appendixA}

Here we describe the procedure that produced the constraints on 
the quadratic potential shown in Fig.~\ref{plot:m2phi2map}
by using slow roll relations to map constraints on the empirical parameters
$\thetaemp=\{\ln A_s,n_s,r\}$ to constraints on the potential parameters 
$\thetapot=\{\Npiv,\log m^2\}$.
For the potential $V=m^2\phi^2/2$, the slow roll approximation gives
\begin{eqnarray}
m^2 &\approx& 24\pi^2 A_s \epsilon_V^2, \label{eq:sr1}\\
\Npiv &\approx& \frac{1}{2}(\epsilon_V^{-1}-1), \label{eq:sr2}\\
\epsilon_V &\approx& \frac{1}{4}(1-n_s) \approx \frac{r}{16}, \label{eq:sr3}
\end{eqnarray}
where $\epsilon_V \equiv (\Mpl^2/2)(V_{,\phi}/V)^2$.

We begin with constraints on $\thetaemp$ from a standard MCMC analysis of 
WMAP7 data, which treats $A_s$, $n_s$, and $r$ as free, independent parameters.
We then postprocess the empirical parameter chains to express the 
constraints in terms of $\thetapot$ by first imposing 
new priors on $\thetaemp$, and then using Eqs.~(\ref{eq:sr1})--(\ref{eq:sr3}) 
to compute the values of $\thetapot$ for each MCMC sample.

The first prior applied to $\thetaemp$ enforces the slow roll relation 
between $n_s$ and $r$ for the quadratic potential, $r\approx 4(1-n_s)$, 
by selecting only those points from the chains that satisfy 
$|1-n_s-r/4|<\delta$. The parameter $\delta$ can be adjusted either 
to match the slow roll relation more closely in the limit $\delta\to 0$, 
or to retain more MCMC points for better sampling of the likelihood 
function (and thus smoother contours) by choosing a larger value of 
$\delta$; for Fig.~\ref{plot:m2phi2map}
we have chosen $\delta=0.005$. Note that the allowed values of $A_s$ are not 
restricted by specializing to the case of the quadratic potential.

We additionally exclude points for which $\Npiv$ is large enough that 
$\rhorh$ is greater than $(3/2) \vend$, 
thus violating energy conservation. (The factor of $3/2$ arises because the 
energy density at the end of inflation contains a significant contribution 
from the kinetic energy.) That is, we require $\Npiv\leq \Npivirh$, where
\begin{equation}
\Npivirh=\ln(\Hpiv/\kpiv)-71.1-\ln(\vend^{1/4}/\Mpl),
\end{equation}
with $\vend\approx m^2 \Mpl^2$ for the quadratic potential.

For the MCMC samples that remain after applying the cuts on $n_s$, $r$, 
and $\Npiv$, we multiply the likelihood of each sample by $r^{-2}$ 
to approximate flat priors on 
$\thetapot$ instead of the original flat priors on $\thetaemp$. 

After including each of these priors and mapping the MCMC parameters 
from $\thetaemp$ to $\thetapot$ using Eqs.~(\ref{eq:sr1})--(\ref{eq:sr3}), 
the resulting constraints shown in Fig.~\ref{plot:m2phi2map} match 
the more accurate constraints from \ModeCode\ (Sec.~\ref{sec:monomial})
reasonably well, albeit with noisy contours due to poor sampling 
along the line $r\approx 4(1-n_s)$. The remaining systematic offset 
between the two sets of constraints is consistent with the effects of 
second order corrections to the slow roll approximation.

\mbox{}

%%%%%%%%%%%%%%%%%%%%%%%%%%%%%%%%%%%%%%%%%
\bibpreamble{\vspace{-1cm}}
\bibliography{bibliography}
%%%%%%%%%%%%%%%%%%%%%%%%%%%%%%%%%%%%%%%%%

\end{document}